\documentclass[namedreferences]{solarphysics}
\usepackage[optionalrh,solaenum]{spr-sola-addons} 
\usepackage{graphicx}        
\usepackage{color}           
\usepackage{url}             

\newcommand{\be}{\begin{equation}}
\newcommand{\ee}{\end{equation}}
\newcommand{\bex}{\begin{equation}\notag}
\newcommand{\eex}{\end{equation}\notag}
\newcommand{\bea}{\begin{eqnarray}}
\newcommand{\eea}{\end{eqnarray}}
\newcommand{\beax}{\begin{eqnarray*}}
\newcommand{\eeax}{\end{eqnarray*}}
\newcommand{\ba}{\begin{array}}
\newcommand{\ea}{\end{array}}

\begin{document}
\begin{article}

\begin{opening}

\title{Radiative Cooling in MHD Models of the Quiet Sun
Convection Zone and Corona}

\author{W.P. \surname{Abbett}$^{1}$ \sep
        G.H. \surname{Fisher}$^{1}$}
\runningauthor{W.P. Abbett, G.H. Fisher}
\runningtitle{Quiet Sun Convection-Zone-to-Corona Models}

\institute{$^{1}$ Space Sciences Laboratory, University of California, 
   Berkeley, CA, USA\\
   email: \url{abbett@ssl.berkeley.edu}, 
          \url{fisher@ssl.berkeley.edu}}

\begin{abstract}
We present a series of numerical simulations of the quiet Sun plasma
threaded by magnetic fields that extend from the upper convection zone 
into the low corona.  We discuss an efficient, simplified approximation to 
the physics of optically thick radiative transport through the surface 
layers, and investigate the effects of convective turbulence on the 
magnetic structure of the Sun's atmosphere in an initially unipolar 
(open field) region.  We find that the net Poynting flux below 
the surface is on average directed toward the interior, while in the 
photosphere and chromosphere the net flow of electromagnetic energy 
is outward into the solar corona.  Overturning convective motions between 
these layers driven by rapid radiative cooling appears to be the source of 
energy for the oppositely directed fluxes of electromagnetic energy.
\end{abstract}

\end{opening}

\section{Introduction}
\label{intro} 

To understand the physics of solar activity, we must understand 
the magnetic and energetic connection between the Sun's convective envelope 
and corona.  The magnetic fields that mediate or energize most, if not all,
solar activity are generated below the visible surface within the 
turbulent convection zone.  Yet most of what we can directly measure
originates from the solar atmosphere, where physical conditions are 
fundamentally different from that of the interior.  While helioseismic 
inversions provide an invaluable window into the physics of the Sun's 
interior, understanding the physical connection between subsurface
features and those observed in the solar atmosphere requires a 
realistic forward model.
 
But what level of realism in a numerical model is necessary 
to describe the complex magnetic connectivity and energetics of
the solar atmosphere lying between the visible surface and the 
corona?  It is of great benefit, for example, to formulate a simple, 
well-defined problem, and set up an idealized numerical experiment that 
sheds light on the relevant physical processes in an otherwise complex 
system.  In this way, important progress has been made in our understanding
of the physics of magnetic flux emergence in highly stratified model
atmospheres (see, {\it e.g.}, \opencite{Manchester2004}; 
\opencite{Murray2006}; \opencite{Magara2006}; \opencite{Galsgaard2007};
\opencite{Fan2009b}; \opencite{Archontis2010}).

Yet the observed evolution of the photospheric magnetic field is often
far more complex, particularly in and around CME- and flare-producing active
regions.  It is difficult to set up a simple magnetic and energetic 
configuration and an associated physics-based photospheric boundary 
condition that can initialize a simulation of the solar atmosphere and 
faithfully mimic the coronal evolution of a complex active region.  If 
we wish to perform first-principles quantitative studies of phenomena
such as eruptive events, the energization of 
the solar wind, active region decay, the transport of magnetic free 
energy and helicity into the solar atmosphere, and the physics of coronal 
heating, it is essential to
evolve a turbulent model convection zone and corona within a single, 
large-scale computational domain.

To achieve this, we must accommodate the fundamental energetics of the 
system while still retaining the ability to study the interplay between 
large and small-scale magnetic structures that evolve over different 
timescales.  Clearly, radiative transport plays a critical 
role in the energy balance of the atmospheric layers that bridge the gap 
between the visible surface and corona.  Surface cooling drives convection, 
and convective turbulence both generates magnetic field and mediates the 
flux of magnetic energy that enters the solar atmosphere.
Yet the physics of radiative transport can be computationally expensive
to treat realistically, even in the context of small-scale domains that 
do not include the convection zone and corona within a single 
computational volume.  For example, energetically important transitions 
in the solar chromosphere are often decoupled from the local thermodynamic 
state of the plasma (a state of non-local thermodynamic equilibrium, 
or non-LTE) suggesting that a truly realistic numerical model must also 
couple the macroscopic radiative transfer and level population equations 
to the system of conservation equations (see, {\it e.g.},
\opencite{McClymont1983}; \opencite{Fisher1985}; \opencite{Carlsson1992};
\opencite{Abbett1999}; \opencite{Allred2005}).  To complicate matters 
further, non-thermal physics, and the physics of ion--neutral drag may 
substantially affect the energy balance of the chromosphere 
\cite{Krasnoselskikh2010}.

While it remains impractical to perform large-scale, 3D, non-LTE radiative
MHD calculations without employing substantial approximations to make
the system tractable, it is now common practice to realistically treat
optically thick surface cooling in the upper convection zone and 
photosphere in LTE (a good approximation in these layers). There 
are many examples of thin-layer, high-resolution calculations 
that incorporate solutions to the non-gray radiative transfer equation in 
Cartesian domains that include the upper convection zone and extend into the 
low chromosphere (\opencite{Bercik2002}; \opencite{Stein2006};
\opencite{Georgobiani2007}; \opencite{Rempel2009a}; \opencite{Cheung2010}). 
In addition, calculations that realistically treat radiative transfer have 
been applied to simulations of solar granulation in relatively small-scale 
domains that also include a transition region and corona 
(\opencite{Martinez2008}; \opencite{Martinez2009}; \opencite{Carlsson2010}).

Our goal, however, is to expand the size of such computational domains
to active region or even global spatial scales while still retaining as
realistic a thermodynamic environment as is feasible.  
Thus, we strive to develop the simplest model possible that allows us 
to capture the essential physics of the convection-zone-to-corona 
system while still maintaining the computational efficiency of models
in which optically thick radiative cooling is treated in a parameterized 
fashion ({\it e.g.}, \opencite{Abbett2007}; \opencite{Fang2010}).  In this 
way, we hope to make practical the performance of physics-based, first 
principles simulations, allowing for quantitative, parameter-space studies of 
processes such as filament formation, active region emergence and decay, and 
flare and CME initiation.

To simultaneously evolve a realistic model convection zone and 
corona at any spatial scale presents a number of daunting 
challenges.  The upper convection zone and low solar atmosphere 
are highly stratified --- average thermodynamic quantities change by many 
orders of magnitude as the domain transitions from a relatively cool, 
turbulent regime below the visible surface, to a hot, magnetically-dominated 
and shock-dominated regime high in the corona.  The physics of 
the gas transitions from a high-$\beta$ plasma where the magnetic field
is advected by the gas (away from strong active region complexes) to 
a low-$\beta$ regime where the gas is constrained to move along magnetic
field lines.  In addition, the radiation field transitions 
from being optically thick to optically thin.  Temporal and spatial 
scales are highly disparate.  Large concentrations of magnetic flux 
are compressed within intergranular lanes and evolve at convective 
turnover timescales, while large coronal loops form and persist for days
as active regions emerge and evolve over a course of many months.  In 
addition, the large-scale magnetic structure of the corona can change in 
a fraction of a second, as small-scale localized magnetic reconnection 
suddenly reorganizes the large-scale field, often triggering eruptive events 
along the way. 
 
The corona presents particular challenges.  It is well known
that in order to accurately reflect the thermodynamics of this region,
a model should include the effects of electron heat conduction along
magnetic field lines and radiative cooling in the optically thin ``coronal
approximation''.  In addition, some physics-based ({\it e.g.}, Joule 
heating) or empirically based 
source of coronal heating must be present (often introduced at the lower 
photospheric boundary) if the model corona is to remain hot.  But to 
generate a realistic magnetic carpet, and to study the interaction of 
granular convection with coronal structures, requires there to be a turbulent
model convection zone, and therefore some form of optically thick surface 
cooling.  

In \inlinecite{Abbett2007} we introduced this physics into a 3D MHD 
convection-zone-to-corona model in the simplest, most computationally efficient 
way possible --- we simply ignored the optically thick radiative transfer 
equation entirely, and instead used a parameterized Newton cooling function 
carefully calibrated against smaller-scale, more realistic radiative-MHD 
models of magneto-convection where the frequency-dependent LTE transfer 
equation was solved along with the MHD system \cite{Bercik2002}.

This approach has been successful in studying the structure
of quiet-Sun magnetic fields and active region flux emergence 
(\opencite{Abbett2007}; \opencite{Fang2010}; \opencite{Fang2010b}).
Yet this treatment, while computationally efficient, has a number of 
limitations.  Its principle drawback is that it is ultimately \emph{ad hoc} 
and requires other, more realistic simulations as a basis for calibration 
in order to get meaningful results.  The simplified cooling is imposed at 
a particular height or over a range of gas density, and is not generated 
in a physical way as a function of optical depth.  To address these 
limitations, we build upon a technique introduced by \inlinecite{Abbett2010}, 
and in Section \ref{radtrans} derive a simple, flux-conservative approximation 
to optically thick cooling that is based on the gray radiative transfer 
equation in LTE.  We then incorporate a form of this efficient, 
physics-based approximation into the {\sf RADMHD} convection-zone-to-corona 
model of \inlinecite{Abbett2007}, which we briefly describe in 
Section \ref{method}.  In Section \ref{results} we present new models of 
an open-field coronal hole region, and study the transport of magnetic 
energy from below the surface into the corona.  Finally, in 
Section \ref{conclusion} we summarize our results.

\section{Numerical Methodology}
\label{method}

The parallel code {\sf RADMHD} solves the following MHD conservation equations 
semi-implicitly on a three-dimensional Cartesian mesh:
\begin{equation}
\frac{\partial\rho}{\partial t}+{\bf\nabla\cdot}\left(\rho
  {\bf u}\right)=0,\label{cont}
\end{equation}
\begin{equation}
\frac{\partial\rho{\bf u}}{\partial t}+{\bf\nabla\cdot}\left[\rho
  {\bf uu}+\left(p+\frac{B^2}{8\pi}\right){\bf I}-\frac{{\bf BB}}{4\pi}
  -{\bf\Pi}\right]=\rho{\bf g},\label{mom}
\end{equation}
\begin{equation}
\frac{\partial{\bf B}}{\partial t}+{\bf\nabla\cdot}\left({\bf uB}
  -{\bf Bu}\right)=-{\bf\nabla\times}\left(\eta{\bf\nabla\times B}
  \right),
\end{equation}
\begin{equation}
\frac{\partial e}{\partial t}+{\bf\nabla\cdot}\left(e{\bf u}
  \right)=-p{\bf\nabla\cdot \,}{\bf u}+\frac{\eta}{4\pi}
  |{\bf\nabla\times B}|^2+\Phi+Q,\label{energy} .
\end{equation}

The components of the state vector have the usual definitions:
$\rho$, ${\bf u}$, $e$, $p$, ${\bf B}$, and ${\bf g}$
denote the gas density, velocity, internal energy per unit volume,
gas pressure, magnetic field, and gravitational acceleration
respectively. Here, we assume Gaussian units.  The viscous
stress tensor is assumed to be of the form $\Pi_{ij}=2\rho\nu
[D_{ij}-1/3({\bf\nabla\cdot u})\delta_{ij}]$, where $D_{ij}=
1/2(\partial u_i/\partial x_j+\partial u_j/\partial x_i)$ and 
$\delta_{ij}$ denotes the Kronecker delta function.  The function 
$\Phi=\sum_{i,j}\Pi_{ij}D_{ij}$ represents the rate of energy
dissipation through viscous diffusion, and $\nu$ and $\eta$
refer to the coefficients of kinematic viscosity and magnetic 
diffusivity, respectively.  These coefficients are assumed constant, 
and are set to values that correspond to the grid-scale viscous
and resistive dissipation.  The source term $Q$ includes important 
energy sources and sinks such as radiative cooling, the divergence 
of the electron heat flux (in the portion of the domain representing 
the model transition region and corona), and any desired 
empirically based coronal heating function.  A complete discussion 
of the components of this energy source term is provided by 
\inlinecite{Abbett2007}.  The system is closed with a non-ideal 
equation of state, using tabular data provided by the {\sf OPAL} project 
\cite{Rogers2000}. In this article, the portion of the domain
corresponding to the corona is heated by the empirically based
coronal heating function described in \inlinecite{Abbett2007},
and the effects of Joule dissipation within this region are 
ignored.

The semi-implicit numerical scheme is parallelized on a domain-decomposed 
mesh, and the core technique is based on operator splitting 
with a high-order Crank--Nicholson temporal discretization.  We treat
the electron thermal conduction, viscous and Joule dissipation, and
radiative losses implicitly using a Jacobian-Free Newton-Krylov ({\sf JFNK})
solver, and require that the remainder of the system be treated explicitly 
using the Central Weighted Essentially Non-Oscillatory ({\sf CWENO}) method of 
(\opencite{Kurganov2000}; \opencite{Balbas2006}).  In this way, we remain 
Courant limited by the magnetosonic wavespeed, and can follow the dynamics 
of the system in a reliable way (we may choose to relax this constraint 
when evolving active region magnetic fields over longer timescales).  Any 
local divergence error introduced into the magnetic field as a result of the 
{\sf CWENO} central scheme is dissipated by adding 
an additional artificial source 
term proportional to ${\bf\nabla(\nabla\cdot B)}$ to the induction
equation.  A detailed description of the numerical methodology employed 
by {\sf RADMHD} can be found in Section 2 of \inlinecite{Abbett2007}.

A number of enhancements and improvements have been incorporated
into the {\sf RADMHD} source code since its initial release in 2007.  Most
improvements are in the form of improved performance and robustness, 
better MPI load balancing and scaling, and other enhancements in
the code's speed and efficiency.  Among the enhancements are: 
{\it i}) a simplified and improved table inversion and 
interpolation algorithm that is necessary to incorporate the {\sf OPAL}
data into the code's non-ideal equation of state and the {\sf CHIANTI} data
\cite{Young2003} into the code's treatment of optically thin radiative 
cooling; {\it ii}) a new adaptive error algorithm in the {\sf GMRES} 
(Generalized Minimum RESidual) substep of the {\sf JFNK} solver that greatly 
improves convergence rates; {\it iii}) a more robust, global non-linear 
{\sf CWENO} weighting scheme in the explicit substep of {\sf RADMHD}; 
and {\it iv}) an option to evolve $\log \rho$ rather that $\rho$ itself 
via the following rewrite of Equation (\ref{cont}):
\begin{equation}
\frac{\partial\ln\rho}{\partial t}+{\bf\nabla\cdot}\left(
  {\bf u}\ln\rho\right)=\left(\ln\rho-1\right){\bf\nabla\cdot}{\bf u} .
\end{equation}
Since the model atmosphere is highly stratified, this is often useful 
as a means of making the code more robust, while at the same time retaining
the desired shock-capture characteristics of the numerical scheme.  
More details on these and other algorithmic improvements will be
provided in a technical document under preparation for inclusion
with the next release of the code.

In essence, however, the core numerical methods of {\sf RADMHD} remain the 
same as that presented in \inlinecite{Abbett2007}.  In this article, we 
focus on the portion of the energy source term $Q$ of Equation (\ref{energy})
that contains the approximation for optically thick radiative cooling.

\section{An Approximate Treatment of Optically Thick Cooling}
\label{radtrans}

Radiative cooling drives surface convection and is a crucial 
contributor to the energy balance in the region of the solar atmosphere 
bridging the convection zone and corona.  Yet a full frequency-dependent
solution to the LTE radiative transfer equation can be computationally 
expensive for large-scale convection-zone-to-corona calculations, 
particularly for active region or filament models where timescales are 
such that the radiative cooling must be updated at intervals close to the 
MHD CFL limit.  Here, we build upon the approach introduced by
\inlinecite{Abbett2010}, and derive an approximate, frequency-integrated
expression for optically thick radiative cooling that is based on the 
gray transfer equation in LTE.  We begin by considering the net cooling
rate for a volume of plasma at a particular location in the solar 
atmosphere:
\begin{equation}
  R=\!\int\!\mathrm{d}\Omega\!\int\!\mathrm{d}\nu
    \left(\eta_\nu-\kappa_\nu I_\nu\right) .
\end{equation}
Here, $\Omega$ represents solid angle, and $\nu$ the frequency.  The 
subscript $\nu$ indicates that the emissivity, opacity, and specific 
intensity ($\eta_\nu$, $\kappa_\nu$, and $I_\nu$ respectively) depend 
on frequency.  If we define the source function $S_\nu$ as the ratio of the
emissivity to opacity, and rearrange the order of integration, we can
recast the net cooling rate in the following form 
\begin{equation}
  R=\!\int\!\mathrm{d}\nu\,\kappa_\nu\!\int\!\mathrm{d}\Omega
    \left(S_\nu-I_\nu\right) .
  \label{eqncool}
\end{equation}
We define the mean intensity as $J_\nu\equiv(1/4\pi)\int\!\mathrm{d}
\Omega\,I_\nu$ and note that the source function is independent of direction.  
This allows Equation (\ref{eqncool}) to be expressed as 
\begin{equation}
\label{netcool}
  R=4\pi\!\int\!\mathrm{d}\nu\,\kappa_\nu\left(S_\nu-J_\nu\right) .
\end{equation}
If we now assume a locally plane-parallel geometry, the formal solution
for the  specific intensity can be written as ({\it e.g.},
\opencite{Mihalas1978})
\begin{equation}
  I_\nu (\tau_\nu ,\mu)=\int_{0}^{\infty} \mathrm{d}\tau^{\prime} 
    \frac{\mathrm{e}^{-|\tau_\nu -\tau^{\prime}|/|\mu|}}{|\mu |}
    S_\nu(\tau^{\prime}) ,
\end{equation}
where $\mu$ refers to the cosine angle and $\tau_\nu$ to the
frequency-dependent optical depth.  We can now recast the expression for 
the mean intensity in terms of an integral over optical depth and cosine 
angle,
\begin{equation}
  J_\nu(\tau_\nu) =\frac{1}{2}\int_{0}^{\infty} \mathrm{d}\tau^{\prime}
    S_\nu (\tau^{\prime})\int_0^1 \mathrm{d}\,|\mu |
    \frac{\mathrm{e}^{-|\tau_\nu-\tau^{\prime}|/|\mu|}}{|\mu |} ,
\end{equation}
This allows the integral over $\mu$ to be evaluated and expressed in terms
of an exponential integral function,
\begin{equation}
\label{jmean}
  J_\nu(\tau_\nu) =\frac{1}{2}\int_{0}^{\infty}\mathrm{d}\tau^{\prime}
    S_\nu(\tau^{\prime})E_1(|\tau_\nu-\tau^{\prime}|) .
\end{equation}

Up to now, no approximation other than an assumption of a locally
plane parallel geometry has been made.  We now follow the analysis
of \inlinecite{Abbett2010} and note that the first exponential integral function 
[$E_1(|\tau_\nu-\tau^{\prime}|)$] in Equation (\ref{jmean}) is singular 
when $\tau^{\prime}=\tau_\nu$, and that this singularity is integrable.  
Since $E_1$ is peaked around $\tau_\nu$, contributions from 
$S_\nu(\tau^\prime)$ will be centered around $S_\nu(\tau_\nu)$.  
Thus, to lowest order we can approximate the mean intensity by
\begin{equation}
  J_\nu(\tau_\nu) \approx \frac{1}{2}S_\nu(\tau_\nu)\int_{0}^{\infty}
    \mathrm{d}\tau^{\prime}E_1(|\tau_\nu-\tau^{\prime}|) .
\end{equation}
The integral over optical depth is now easily evaluated, and the result
is expressed in terms of the second exponential integral function [$E_2$]: 
\begin{equation}
  J_\nu(\tau_\nu) \approx S_\nu(\tau_\nu)\left( 1-\frac{E_2(\tau_\nu)}{2}
     \right) .
\end{equation}
We now rearrange the terms in the above equation, and substitute
$1-J_\nu(\tau_\nu)/S_\nu(\tau_\nu)\\
\approx E_2(\tau_\nu)/2$ into 
Equation (\ref{netcool}), to arrive at an approximation for the net 
cooling rate,
\begin{equation}
  R\approx 2\pi\!\int\!\mathrm{d}\nu\,\kappa_\nu S_\nu E_2(\tau_\nu) .
\end{equation}
If we further assume LTE, the source function can be expressed as 
the Planck function [$B_\nu(T)$] coupling the cooling rate to the local 
temperature of the plasma [$T$]:
\begin{equation}
  \label{cooling2cool}
  R\approx 2\pi\!\int\!\mathrm{d}\nu\,\kappa_\nu B_\nu (T) E_2(\tau_\nu) .
\end{equation}

We now integrate Equation (\ref{cooling2cool})
over frequency.  Since $E_2(\tau_{\nu})$ is bounded below by zero and 
above by unity, the integral in Equation (\ref{cooling2cool}) 
obeys this set of inequalities:
\begin{equation}
  2\bar\kappa\sigma T^4 > 2\pi \int \mathrm{d}\nu\kappa_{\nu}B_{\nu}(T)
    E_2(\tau_{\nu}) > 0\ ,
\label{equation:inequalities}
\end{equation}
where $\bar\kappa$ is the Planck-weighted mean opacity.

Because of the range of the $E_2(x)$ function, we can use this inequality
to write the integral in Equation (\ref{cooling2cool}) in the 
form $R(\bar\tau)=2\bar\kappa C(\bar\tau)\sigma T^4 E_2(\alpha
(\bar\tau)\bar\tau)$, where in general, $\alpha$ is a positive, unknown 
function of mean optical depth $\bar\tau$ ($\mathrm{d}\bar\tau\equiv 
-\bar\kappa \mathrm{d}z$), 
and $C(\bar \tau)$ is an unknown, $\bar\tau$-dependent normalization 
constant.  However, since we expect a close relationship between the mean 
optical depth $\bar\tau$ and the local mean opacity $\bar\kappa$, we 
therefore make the ansatz that $\alpha$ is a constant, but with an unknown 
value.  The expression for $R$ can then be written
\begin{equation}
  \label{cooling2}
  R\approx 2C\,\overline{\kappa}\sigma T^{\,4} E_2(\alpha\overline{\tau}) ,
\end{equation}
where $C$ now represents a $\bar\tau$-independent normalization constant
of integration.

To determine the normalization constant $C$, we integrate our cooling 
function from zero to infinity in optical depth over an isothermal 
slab to obtain the total radiative flux.  The resulting expression must 
be equal to the known result $F_{\mathrm{tot}}=\sigma T^4$, 
thus requiring $C=\alpha$.
To calibrate $\alpha$, we compare the cooling rate as a function of depth
in test models using this approximation against more realistic models
of magnetoconvection where the frequency-dependent transfer equation is
solved in detail \cite{Bercik2002}.  We conclude that the best-fit
value is $\alpha=1$ (see Figure 1 of \opencite{Abbett2010}).  This implies 
that optical depth in highly stratified atmospheres is dominated
by the local opacity.

With the parameter $\alpha$ specified, we arrive at an approximation
for optically thick surface cooling, 
\begin{equation}
\label{cooling}
  R\approx 2\,\overline{\kappa}\sigma T^{\,4} E_2(\overline{\tau}) .
\end{equation}
This expression can be efficiently evaluated at each iteration of an 
MHD calculation, and we have implemented this volumetric cooling rate
as a part of the cell-centered source term $Q$ in Equation (\ref{energy})
(the cooling rate $R$ being a negative heating rate $Q$).

It is possible for the computational grid to be of 
sufficient resolution to resolve the local pressure scale heights
of a highly stratified model atmosphere while at the same time
being poorly resolved in optical depth.  This has the potential
to lead to numerical error in the calculation of the local cooling
rate such that the total radiative flux may not be conserved.  We
therefore consider a flux conservative formulation similar to the
constrained transport schemes common to many MHD codes (see 
\opencite{Stone1992}).

We begin by defining a frequency-independent discretized, optical 
depth for each iteration where the MHD state variables are updated.  
Since our simple approximation is based on the assumption of a locally 
plane-parallel geometry, and we are neglecting (for now) the effects 
of sideways transport, all that is required is an integration along 
the vertical direction ({\it i.e.}, in the direction of the gravitational 
acceleration).  Our discretized expression takes the form  
\begin{equation}
  \tau_{i,j,k-1/2}\equiv\sum_{n=k_{top}}^{k}\overline{
     \kappa}_{i,j,k}\left(z_{n+1/2}-z_{n-1/2}\right)
\end{equation}
Here, the grid coordinates $i,j,k$ are defined at cell centers 
(consistent with the centralized numerical scheme implemented in {\sf RADMHD}), 
and the optical depth is defined at face centers of the mesh cell's control 
volume perpendicular to the $z$-direction (we now drop the overbar notation,
since the above definition makes it clear that $\tau$ is a frequency-averaged
quantity).  We use tabular Planck-weighted
mean opacities ($\overline{\kappa}_{i,j,k}$) provided by the opacity 
project \cite{Seaton2005}.  The coordinate $k_{top}$ refers to the first 
ghost cell of the upper coronal boundary of the simulation domain, though 
in practice it is set to an interior cell bounding the portion of the domain 
that represents the optically thin corona.  Either way, it is presumed that 
$\tau_{i,j,k_{top}-1/2}=0$.  Note that optical depth increases 
inward into the atmosphere in the opposite sense of the height $z$, which 
is defined to increase outward from the interior toward the visible surface 
({\it i.e.}, $\mathrm{d}\tau\equiv-\overline{\kappa} \mathrm{d}z$).   

The radiative cooling of Equation (\ref{cooling}) can be expressed in terms
of a divergence of a radiative flux.  Our treatment of radiative
transfer assumes a locally plane-parallel geometry, thus we need only 
consider the radiative flux at the faces of control volumes normal to
the vertical direction.  This implies that any horizontal divergence of 
the radiative flux is assumed negligible when compared to gradients in 
the vertical direction.  The physical justification for this simplification 
is that changes in emissivity and opacity are generally much greater in the 
vertical direction of a highly stratified atmosphere than those expected in 
the transverse direction.  This assumption will likely not be valid at the 
edges of sunspots where the lateral emissivity and opacity gradients are 
expected to be large. Given this simplification, the divergence of the 
radiative flux can be expressed as 
\begin{equation}
   \frac{\partial F}{\partial z}=2\overline{\kappa}\sigma T^{\,4} 
      E_2(\tau) .
\end{equation}
We now cast this
expression in terms of optical depth $\mathrm{d}\tau\equiv 
-\overline{\kappa}\mathrm{d}z$:
\begin{equation}
\label{flux}
   \frac{\partial F}{\partial \tau}=-2\sigma T^{\,4} 
      E_2(\tau) ,
\end{equation}
and note that this equation is of the form $Q(\tau)=-f(\tau)E_2(\tau)$
with $f(\tau)\equiv 2\sigma T^{\,4}$.  We now approximate $f(\tau)$ with
a Taylor-series expansion centered about $\tau_k$ accurate up to second 
order, and reorder the terms so that the expression is of the form 
$f(\tau)=A+B\tau$:
\begin{equation}
  f(\tau)=\left[f(\tau_k)-\tau_k f^{\prime}(\tau_k)\right]
    +\tau f^{\prime}(\tau_k) .
\end{equation}
Here, $f(\tau_k)=2\sigma T^{\,4}(\tau_k)$ refers to the function $f(\tau)$
evaluated at cell center coordinate $(i,j,k)$, $f^{\prime}(\tau_k)$ refers
to the function's vertical derivative with respect to optical depth evaluated
at the same location, and the constants $A$ and $B$ have the form
$A=[f(\tau_k)-\tau_k f^{\prime}(\tau_k)]$ and $B=f^{\prime}(\tau_k)$.  For 
brevity, we have dropped the $i$ and $j$ subscripts, but note that these 
expressions are valid for all grid cells at a particular height.

To obtain the discretized form of Equation (\ref{flux}), we integrate over
the control volume of the computational cell,
\begin{equation}
\label{fluxintegral}
   F(\tau_{k+1/2})-F(\tau_{k-1/2})=
     -\int_{\tau_{k-1/2}}^{\tau_{k+1/2}}
     (A+B\tau)E_2(\tau) \mathrm{d}\tau.
\end{equation}
This integral can be evaluated using the relation $\mathrm{d}
E_n(\tau)/\mathrm{d}\tau =-E_{n-1}(\tau)$ to obtain
\begin{eqnarray}
\label{fluxform}
  F(\tau_{k+1/2})-F(\tau_{k-1/2})&=&
    A\left[E_3(\tau_{k+1/2})-E_3(\tau_{k-1/2})
    \right]\nonumber\\ 
    &+& \!\! B\left[\tau_{k+1/2}\,E_3(\tau_{k+1/2})
    -\tau_{k-1/2}\,E_3(\tau_{k-1/2})\right.\nonumber\\
    &+&\!\! \left. E_4(\tau_{k+1/2})-E_4(\tau_{k-1/2})
    \right] .
\end{eqnarray}
All that remains is to define a stencil to evaluate $f(\tau_k)$ and
$f^{\prime}(\tau_k)$ in the expressions for $A$ and $B$.  {\sf RADMHD} employs 
a central scheme, and it is desirable to derive a stencil consistent 
with the formalism of the code.  Since the temperature $T$ is obtained 
via a table lookup based on cell-centered values of gas density and 
internal energy per unit volume, we obtain our interpolation stencil by 
expanding elements of the state vector $q(\tau)$ in a second-order
accurate Taylor series about $\tau_k$,
\begin{equation}
\label{statetaylor}
  q(\tau)=a+b(\tau-\tau_k)+\frac{c}{2}(\tau-\tau_k)^2 ,
\end{equation}
and enforce the definition of cell-averaged quantities along the
$z$-coordinate axis [again, the $(i,j)$ dependence is implicitly assumed,
and $\Delta\tau\equiv \tau_{k+1/2}-\tau_{k-1/2}$ is less than zero],
\begin{equation}
\label{conserve}
  q_k=\frac{1}{\Delta \tau}\int_{\tau_{k-1/2}}^{
    \tau_{k+1/2}}q(\tau)\mathrm{d}\tau .
\end{equation}
We then expand about each of the points $\tau_{k+1}$, $\tau_k$, 
and $\tau_{k-1}$; substitute the appropriate form of 
Equation (\ref{statetaylor}) into Equation (\ref{conserve}); then 
perform the integration over each respective control volume.  This 
yields a system of equations whose solution specifies $a$, $b$, and $c$ 
in terms of known cell averages.  The compact stencils are equivalent 
to those of \inlinecite{Abbett2007}, and have the form $a=q_k-(q_{k+1}-
2q_k+q_{k-1})/24$, $b=(q_{k+1}-q_{k-1})/(2\Delta\tau)$, and $c/2=
(q_{k+1}-2q_k+q_{k-1})/(\Delta\tau)^2$.  The second term of $a$ arises
from the integration of the second-order term in the Taylor expansion
of $q(\tau)$, and ensures that the interpolation scheme maintains 
second-order accuracy, and that the following discretized, cell-centered 
forms of $A$ and $B$ have desirable stability properties: 
\begin{equation}
  A=f_k-\frac{1}{24}\left(f_{k+1}-2f_k+f_{k-1}\right)
    -\tau_k\frac{1}{2\Delta\tau}\left(f_{k+1}-f_{k-1}\right)
\end{equation}
\begin{equation}
  B=\frac{1}{2\Delta\tau}\left(f_{k+1}-f_{k-1}\right) .
\end{equation}
Here $f_k=2\sigma T_k^4$, and $\tau_k=\tau_{k-1/2}-\kappa_k(\Delta z/2)$.
With $A$ and $B$ specified, we arrive at an expression for
a flux-conservative approximation to the optically thick radiative
source term,
\begin{eqnarray}
\label{fluxform2}
  Q(\tau_{i,j,k})&=& A\overline{\kappa}_{i,j,k}\left[E_3(\tau_{i,j,k-1/2})
    -E_3(\tau_{i,j,k+1/2})\right]\nonumber\\
    &+&\!\! B\overline{\kappa}_{i,j,k}\left[\tau_{i,j,k-1/2}\,
    E_3(\tau_{i,j,k-1/2})-\tau_{i,j,k+1/2}\,
    E_3(\tau_{i,j,k+1/2})\right.\nonumber\\
    &+&\!\! \left. E_4(\tau_{i,j,k-1/2})
    -E_4(\tau_{i,j,k +1/2})\right] .
\end{eqnarray}
By design, this expression will conserve flux to machine roundoff, 
as can easily be verified by showing that $Q_{k}+Q_{k-1}=F_{k+1/2}-F_{k-3/2}$ 
for all points $(i,j,k)$.  Thus, we have two ways of implementing our 
approximation --- the flux conservative approach of Equation (\ref{fluxform2}), 
and the non-conservative approach obtained by directly evaluating 
Equation (\ref{cooling}) using cell-centered quantities.  The flux-conserving 
method requires additional table lookups each iteration to evaluate the 
exponential integrals, but is helpful in cases where the optical depth scale 
is not particularly well-resolved.  

Our approximation for gray LTE cooling is 
applied only in those regions of the computational domain where such an 
approximation is needed.  Specifically, we apply the approximation over a 
range of optical depths that extend from $\tau=10$ to $\tau=0.1$.  At 
greater optical depths, we use the diffusion approximation (with tabular 
Rosseland mean opacities provided by \opencite{Seaton2005}), and at smaller 
optical depths we use the optically thin approximation (using {\sf CHIANTI} data 
from \opencite{Dere1997} and \opencite{Young2003} to specify the optically 
thin cooling curve) as described in \inlinecite{Abbett2007} and 
\inlinecite{Lundquist2008a}.

The simulations we present in Section \ref{results} use the non-conservative 
technique.  The conservative approach described above was motivated by the 
fact that in some cases, where the model atmosphere is poorly resolved in 
optical depth, the strong cooling prescribed by Equation (\ref{cooling}) can 
be concentrated in a narrow one- or two- zone layer of a model atmosphere.  As 
a practical matter, this required that we enforce a limit on the 
maximum amount of cooling per unit mass allowable in any given grid cell.  
This cooling floor is somewhat artificial, and is not necessary in the flux 
conservative method, which has the effect of spreading the cooling over 
adjoining cells in a more physical way.  Another option would be to have a 
separate grid for optical depth, but we decided against this because of the 
possibility of introducing additional interpolation error that is difficult 
to characterize.  We are currently testing our new flux-conservative scheme, 
and plan to fully implement it in a new radiation subroutine for {\sf RADMHD} 
that also includes the important effects of sideways transport.  We hope 
to report on these efforts in the near future.

\begin{figure}    
   \centerline{\includegraphics[width=1.0\textwidth,clip=]{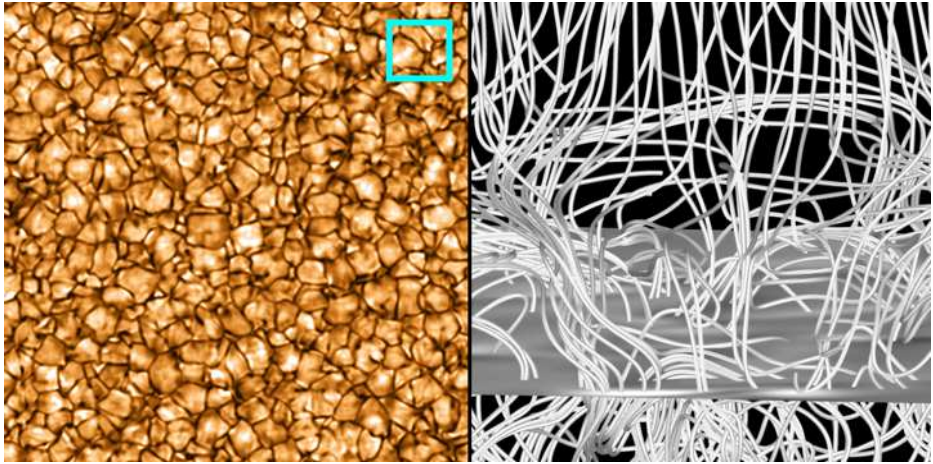}}
   \caption{Left: Temperature at the {\sf RADMHD} model photosphere. Right:
      Magnetic fieldlines threading the low atmosphere over a small 
      sub-domain (the box in the left frame indicates the approximate 
      size of the corresponding sub-domain).  The gray slice indicates
      the average height of the visible surface. The domain spans 
      $24\times 24\times 12$ Mm$\,^3$ at a resolution of $512\times 512\times
      256$.}
   \label{fig1}
\end{figure}

\section{A Model of an Open Flux Region}
\label{results}

We initiate our calculations using the procedure of \inlinecite{Abbett2007}.
Briefly, we begin by relaxing a 1D-symmetric average stratification,
then expand the domain to three dimensions and break the 1D symmetry
by introducing a small, random energy perturbation in the superadiabatically
stratified portion of the computational domain representing the solar
convection zone.  Convective turbulence develops as the simulations
progress, and we allow the model convection zone to dynamically
relax.  We show results from two separate simulations: one was performed
locally using 112 processors of a relatively small Beowulf cluster, and 
the other was performed on NASA's {\sf Discover} supercomputer using 512 
processors.  Both simulations simultaneously evolve a model convection
zone and corona, and each domain has a vertical extent of of 12 Mm, 
with a 2.5 Mm deep model convection zone.  The development model that
was run on the local Beowulf cluster has a domain that spans
$21\times 12\times 12$ Mm$\,^3$ at a resolution of $448 \times 256\times
256$, while the larger run on {\sf Discover} spans $24\times 24\times 12$
Mm$\,^3$ at a relatively high resolution of $512\times 512\times 256$.
In each case, only $\approx 0.66$ percent of the total computational 
effort was expended by the approximate treatment of the radiative transfer
on average within any given MPI subdomain during a given timestep.  On 
the Intel Xeon E5420 CPUs of our local Beowulf cluster, the computing time 
per update of this substep is approximately $0.03$ core-microseconds per 
point.  The simulations presented here should be considered relatively 
small-scale in the context of the capability of the algorithms presented 
--- the code scales well on multiple processors, and once our development 
work is complete we intend to dramatically extend the spatial scale of the 
models.  

The simulations presented here differ from those of 
\inlinecite{Abbett2007} in a fundamental way.  The approximation we
now use for optically thick cooling eliminates all of the \emph{ad-hoc}
calibrated parameters present in the older models.  Specifically,
the height and magnitude of the optically thick radiative source term
is now calculated in a physically self-consistent way based on an
optical-depth scale rather than on an specified density or height
range attenuated by envelope functions ({\it c.f.} Section 2.1.1 of 
\opencite{Abbett2007}).  Once $\alpha$ of Equation (\ref{cooling2}) has 
been calibrated against more realistic models, no further adjustment 
is required, and each atmosphere relaxes to a state determined by the
solution of the system of Equations (\ref{cont})-(\ref{energy}) subject to
imposed boundary conditions.

\begin{figure}    
   \centerline{\includegraphics[width=1.0\textwidth,clip=]{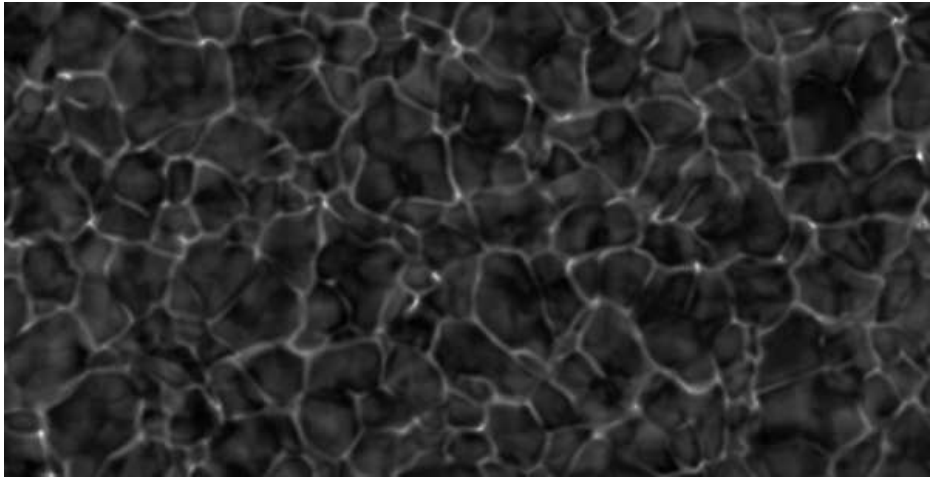}}
   \caption{Temperature in the {\sf RADMHD} low chromosphere showing a
      reverse granulation pattern.  Lighter (darker) colors indicate 
      hotter (cooler) temperatures.  In the models, this occurs because
      the radiative cooling diminishes with height, and the 
      ${\bf p\nabla\cdot v}$ work of converging and diverging flows above 
      the intergranular lanes begins to dominate. The horizontal slice
      spans $21\times 12$ Mm$\,^2$ at a resolution of $448 \times 256$}.
   \label{fig2}
\end{figure}

We apply periodic boundaries in the horizontal directions, and a simple,
somewhat-artificial closed lower boundary.  Specifically, the internal energy
per unit volume within ghost cells adjacent to the domain's lower boundary 
is set such that a temperature gradient is maintained that best matches the 
average stratification at a corresponding height in the \inlinecite{Bercik2002}
magnetoconvection models.  In addition, the ghost cells at the lower boundary 
are specified such that the vertical components of the velocity and magnetic
field and the vertical gradients of the horizontal components of the velocity 
and magnetic field are zero, and such that the gradient of the gas density is 
maintained.  The upper coronal boundary is initially taken to be 
anti-symmetric during the relaxation procedure ({\it i.e.}, ghost zones are set 
such that the vertical component of the velocity and magnetic field vanishes 
at the boundary while all other components of the MHD state vector maintain 
a zero vertical gradient across the boundary interface), then is set to a 
standard zero-gradient boundary condition once magnetic fields are introduced 
({\it i.e.}, ghost cells are set such that all components of the MHD state 
vector maintain a zero vertical gradient across the upper boundary interface).
For the simulations presented here, once the purely hydrodynamic model 
convection zone is relaxed, we introduce a weak 1 G vertically directed 
magnetic field.  This is intended to create an open-flux region, such as 
one might expect within a coronal hole.

As the simulations progress, the convective turbulence acts to stretch 
and amplify the field, and the portion of the domain representing the 
corona begins to heat as a result of the magnetic-field-dependent 
empirically-based coronal-heating source term.  This heating function
is based on the \inlinecite{Pevtsov2003} power-law relationship 
between X-ray luminosity and total unsigned magnetic flux observed 
at the surface (see Equation 12 of \opencite{Abbett2007}).  For this 
study, we are content to rely on empirical heating rather than Joule 
dissipation to energize the model corona since our focus is on the 
transport of magnetic energy into the atmosphere, not the heating 
of the low atmosphere and corona.

Up to the point that magnetic field was introduced into the
simulation domain, the model corona was simply an unphysical, cold, 
nearly evacuated region.  Once the corona heats, 
we activate the implicit electron thermal conduction 
source term.  This builds a corona, but reduces our timestep somewhat since 
the stiffness of the system is increased and convergence rates in the 
{\sf JFNK} substep can become an issue. Thus, it tends to be the last step 
of our relaxation process.  After several additional turnover times, we 
begin our analyses.

\begin{figure}    
   \centerline{\includegraphics[width=1.0\textwidth,clip=]{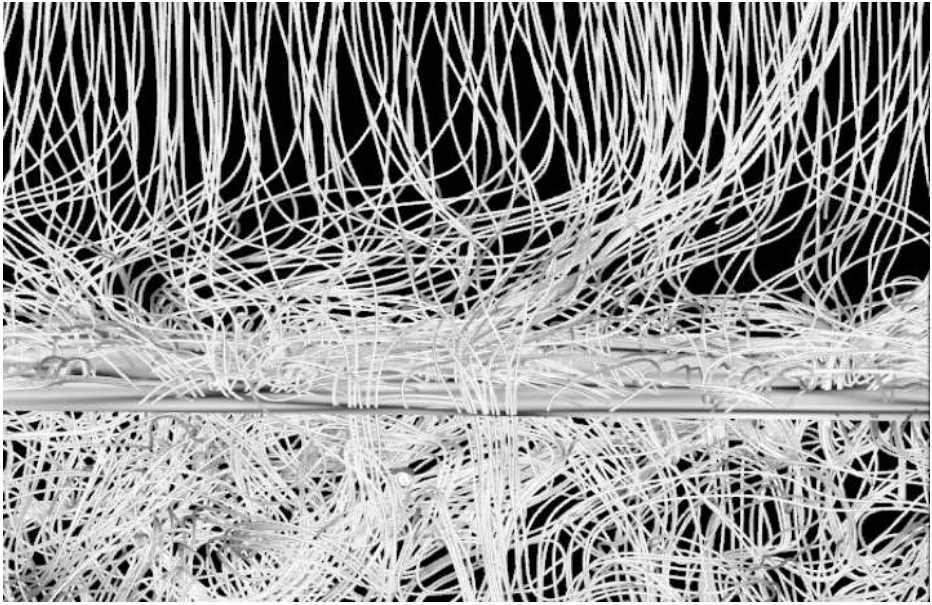}}
   \caption{Magnetic fieldlines threading the low atmosphere over a portion
      of the computational domain.  The gray slice represents the approximate
      position of the model's visible surface.  Horizontally directed magnetic
      fields due to the spreading of canopy-like structures and overturning 
      convective cells permeate the low atmosphere.}
   \label{fig3}
\end{figure}

The left frame of Figure~\ref{fig1} shows the temperature at the visible 
surface ($\tau=1$) of a relaxed convection zone-to-corona model 
using the new formalism of Section \ref{radtrans}.  The granular pattern
and convective turnover times compare well to the realistic magnetoconvection 
models of \inlinecite{Bercik2002}.  This, along with the reverse granulation
pattern shown in Figure~\ref{fig2}, indicates 
that our approximation to optically thick radiative transfer is capturing 
the physics of surface cooling at least well enough to generate and sustain 
solar-like convective features.  

\begin{figure}    
   \centerline{\includegraphics[width=1.0\textwidth,clip=]{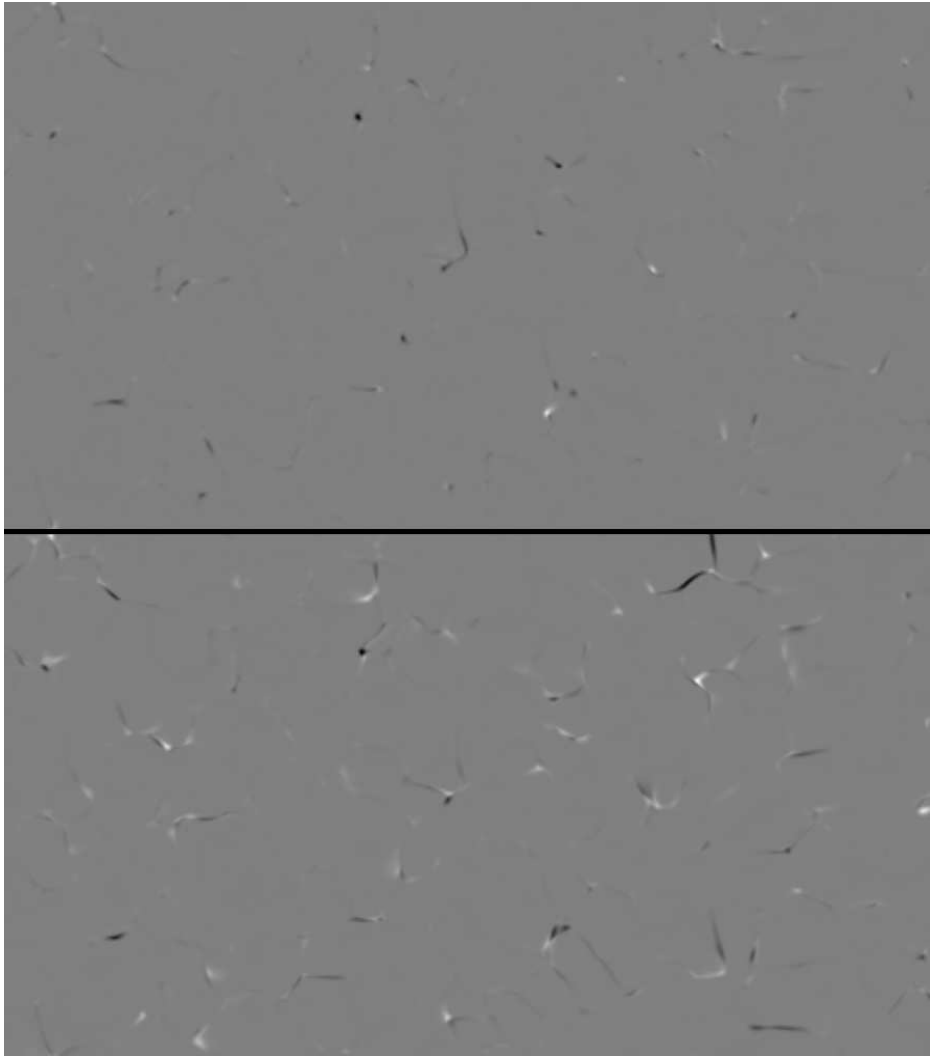}}
   \caption{The vertical component of the Poynting flux along a layer 
      positioned just below optical depth unity (top) and 400 km higher
      near the tops of overturning granules (bottom).  Light colors correspond 
      to outward directed flux (toward the corona); dark colors represent 
      inward-directed flux (toward the convective interior). Each slice
      spans $21\times 12$ Mm$\,^2$. }
   \label{fig4}
\end{figure}

The right frame of Figure~\ref{fig1} shows the complex magnetic structure 
threading a small portion of the simulation domain (as indicated by the 
cyan box in the upper-right corner of the left frame).  Figure~\ref{fig3} 
shows a larger subdomain at a later time, and more clearly illustrates the 
characteristics of the magnetic structure.  In the region where convective 
cells turn over, the average plasma-$\beta$ remains relatively high.  
As a result, much of the magnetic field remains entrained in the plasma, 
turns over, and is recirculated back below the surface.  Thus, at any given 
time, this region is filled with horizontally directed field, and 
that field tends to be less concentrated than is typical of fields entrained
in the vortical downdrafts present at the visible surface and below.  
In addition, the presence of canopy-like structures (where strong 
concentrations of field above intergranular lanes and photospheric 
downdrafts open into the upper atmosphere and spread out like a fan) also 
contribute to the net amount of horizontally directed magnetic field 
threading the atmosphere below the corona.

\begin{figure}    
   \centerline{\includegraphics[width=0.9\textwidth,clip=]{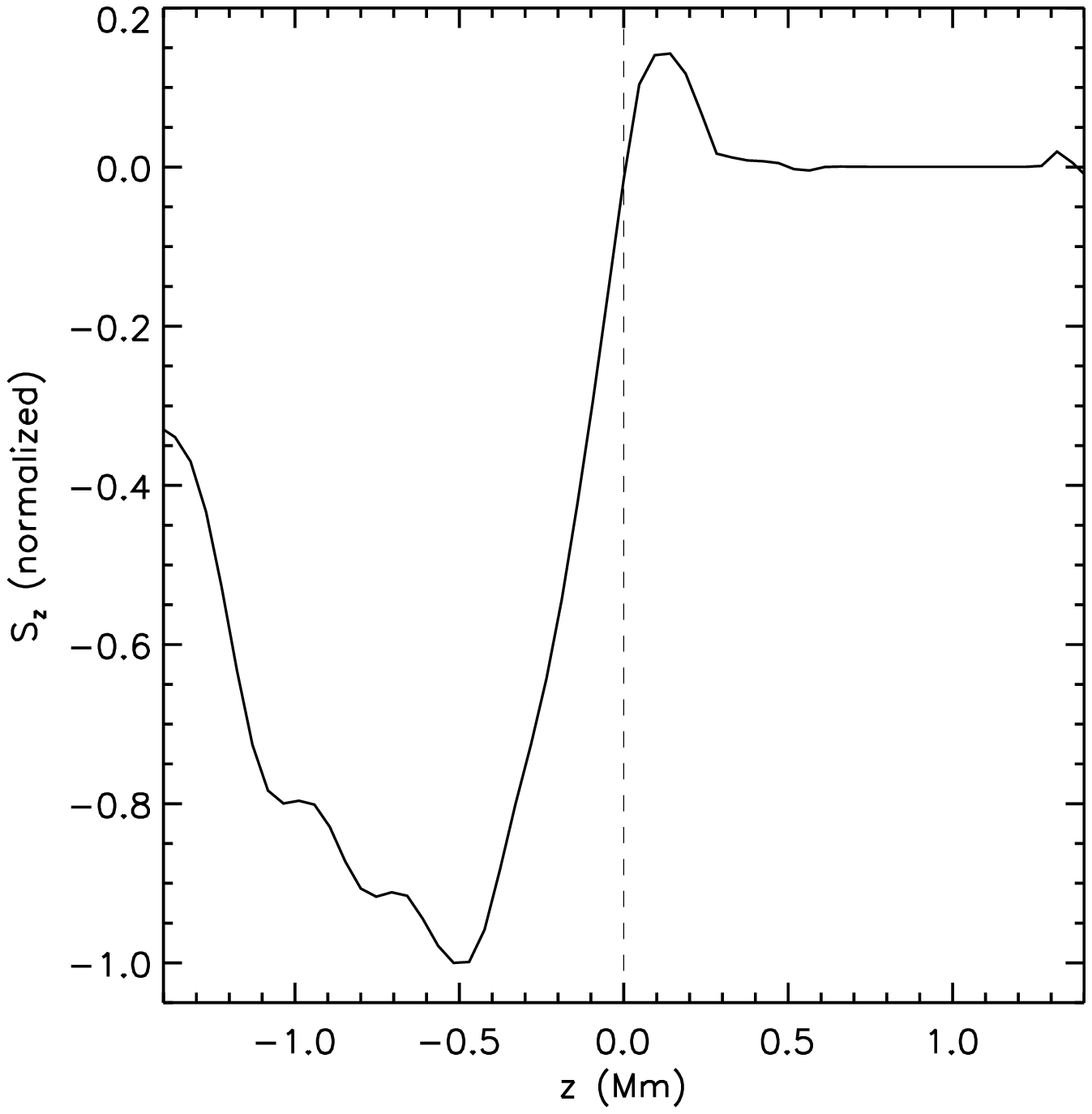}}
   \caption{The normalized net vertical component of the Poynting flux over 
      a portion of the domain centered at the model's photosphere.  The dashed 
      vertical line represents the approximate height of the visible surface.  
      Above the visible surface, electromagnetic energy tends to flow 
      outward toward the corona, while below the surface, energy flows inward 
      toward the convective interior. Above $z\approx 0.5$ (in the model's
      low chromosphere) the Poynting flux tends to remain outwardly directed, 
      but its magnitude is, on average, less than a percent of its maximum 
      value.} 
   \label{fig5}
\end{figure}

The presence of horizontal fields in the atmosphere has consequences for 
the transport of magnetic energy into the upper atmosphere.  Consider the
electromagnetic Poynting flux,
\begin{equation}
{\bf S}=\frac{1}{4\pi}c{\bf E \times B} .
\end{equation}
The vertical component of the Poynting flux is a measure of the amount of
electromagnetic energy flowing into, or out of the solar atmosphere from
below the surface where it is generated.  In Figure~\ref{fig4}, we display 
$S_z$ as a grayscale image at two layers in the model atmosphere 
--- dark shades correspond to a flux of magnetic energy directed toward
the interior, while lighter shades correspond to an outward-directed flux.
The top frame shows the vertical component of Poynting flux along an
$x$--$y$ slice positioned just below the visible surface, 
and the lower frame shows $S_z$ along a slice positioned in
the low atmosphere, $400$ km higher.  Careful examination
of Figure~\ref{fig4} reveals an imbalance in the outward- and inward-directed
flux.  Below the surface, there appears to be a net flow of magnetic energy 
into the convective interior along the strong 
vortical downdrafts contained within intergranular lanes.  Conversely, in
the low atmosphere, the vertical component of the Poynting flux 
appears more diffuse, and there appears to be a net excess of outward directed
flux, particularly within overturning granules.

This is shown more clearly in Figure~\ref{fig5} where 
$S_z$ is integrated over each layer in the computational domain,
and plotted as a normalized quantity as a function of height [$z$].  The
dashed vertical line in the figure represents the average height of the 
visible surface.  What is clear, is that magnetic energy on average
is directed downward into the interior below the visible surface.
It is at the surface and above that the net vertical Poynting flux 
changes sign and becomes outwardly directed.  This suggests that the 
kinetic motion of overturning granules in the model's overshoot layer
provides the source of magnetic energy for the corona, not the deeper 
layers below the optical surface, where magnetic flux and energy 
are being pumped down into the interior along intergranular downflows. 

This can be understood in a fairly straightforward way.  The vertical
component of the Poynting flux can be expressed as 
\begin{equation}
   S_z=\frac{1}{4\pi}\left(c{\bf E}_h {\bf\times}{\bf B}_h\right)
     {\bf\cdot \,\hat{z}} ,
\end{equation}
where ${\bf E}_h$ and ${\bf B}_h$ refer to the horizontal components
of the electric and magnetic field respectively.  If we assume ideal
MHD, then the horizontal component of the electric field can be
written as follows:
\begin{equation}
\label{etrans}
  c{\bf E}_h=-u_z{\bf\hat{z}\times}{\bf B}_h
     -{\bf u}_h{\bf\times}B_z{\bf\hat{z}} .
\end{equation}
Just above the visible surface, the magnetic field remains entrained in 
the fluid as convective cells overturn.  At this height, the strongest 
field concentrations are located near the edges of overturning granules 
as divergent flows from neighboring cells compress the field.  On average, 
there is more of a contribution to the horizontal electric field from 
the second term of Equation (\ref{etrans}), $c{\bf E}_h=-{\bf u}_h{\bf\times}
B_z{\bf\hat{z}}$, since the magnetic field becomes more vertical as 
converging flows compress flux into a relatively small area.  The 
contribution of this term to the vertical component of the Poynting 
flux can be expressed as $4\pi S_z=[(-{\bf u}_h{\bf\times}B_z{\bf\hat{z}})
{\bf \times}{\bf B}_h]{\bf\cdot \, \hat{z}}$.  Simplified, this becomes 
$4\pi S_z=-B_z({\bf B}_h{\bf \cdot} \, {\bf u}_h)$.  

To illustrate the correlation between the horizontal magnetic fields 
[${\bf B}_h$] and the converging surface flows [${\bf u}_h$], consider a 
weak, vertically oriented, untwisted magnetic flux tube that passes 
through the surface.  Suppose it is acted on by a strong converging 
flow in a thin layer at the surface.  If the magnetic field of the tube 
is oriented in the positive $z$ direction, then just above the surface, 
the compression will tilt the fieldlines and create horizontal components 
of the magnetic field in the opposite direction of the converging flow.  
If the magnetic field is oriented in the negative $z$ direction, then 
the horizontal components of the field will be aligned with the flow.  
Either way, $4\pi S_z=-B_z({\bf B}_h {\bf \cdot} \, {\bf u}_h)$ is positive 
above the surface.  Obviously, the dynamics of the model are far more 
complex than this simple thought experiment.  Nevertheless, we do find 
a net positive contribution to the Poynting flux from the second term of 
Equation (\ref{etrans}) along the edges of overturning granules above 
the surface where the field is being compressed. 

Below the photosphere, the situation is quite different.  The strongest
magnetic fields are concentrated within localized vortical downdrafts.  
The asymmetry between these strong downdrafts and the broad upwelling 
plasma in stratified convection is well known, and may provide a mechanism 
whereby magnetic flux can be pumped into the interior (\opencite{Tobias2001}).  
We find more of a contribution to the net Poynting flux from the first 
term of Equation (\ref{etrans}), $4\pi S_z=[(-u_z{\bf\hat{z} \times}{\bf B}_h) 
{\bf \times}{\bf B}_h] {\bf\cdot \, \hat{z}}$.   This can be expressed more 
simply as $4\pi S_z=u_zB_h^2$, and in this form, it is easy to see that a
net downward transport of magnetic flux is consistent with a net 
downward-directed Poynting flux.  This downward-directed flux of 
electromagnetic energy below the surface is consistent with other 
simulations of radiative-magnetoconvection (see, {\it e.g.}, 
\opencite{Vogler2007}).  

The first to recognize this change in direction of the flow of
electromagnetic energy was \inlinecite{Steiner2008} who referred
to the visible surface as ``a separatrix for the vertically directed
Poynting Flux''.   Our results are consistent with their findings,
although we conclude that in the larger domain, the upward-directed net 
flow of magnetic energy tends to arise from the action of the 
compressive flows of overturning convection as magnetic flux
is expelled from cell centers and concentrated into the intergranular
regions.  However, higher in the model
atmosphere as the gas transitions to a low-$\beta$ regime (the
upper chromosphere--transition region boundary), we also see a small
buildup of magnetic flux, for reasons similar to those 
of \inlinecite{Steiner2008}.  Namely, that the dynamic chromosphere
transitions to a stable, subadiabatic, magnetically-dominated regime,
and there is a magnetic reservoir on average as magnetic flux that 
is advected upward enters the stable regime and does not get recirculated
back into the convective interior (similar in some ways to the overshoot 
layer at the base of the convection zone).  This is reflected in the small
peak in Figure~\ref{fig5} at a height of $\approx$ 1.3 Mm above the
visible surface.  Above this transition region interface in the open 
field of the model corona, energy is transported via magnetosonic and 
Alfv\'en waves.  We note that in these simulations, the magnetic field
has yet to fully saturate ({\it i.e.}, there is a small increase in magnetic 
energy over time as magnetic field is stretched and amplified by convective
turbulence).  While this indicates that the atmosphere has yet to fully
relax, this increase in magnetic energy (and any Joule heating below the 
corona) is negligible in comparison to the divergence of the Poynting 
flux and the work done on the magnetic field by convective motions.
 
In some sense, the height at which the transition between outward and 
inward flow of electromagnetic energy takes place is less important than 
the fact that such a transition exists.  What the simulations seem to suggest 
is that in quiescent regions away from particularly strong concentrations 
of magnetic flux, there is not a continuous flow of electromagnetic 
energy from below the surface out into the corona.  Instead, the 
mechanical energy of convection mediates the flow of magnetic 
energy in the relatively high-$\beta$ surface layers where the magnetic 
field remains frozen into the plasma.  Of course, in and around very 
strong concentrations of magnetic flux, the situation is undoubtedly 
quite different. 

\section{Discussion and Conclusions}
\label{conclusion}

We have developed an approximate treatment of optically thick radiative
surface cooling that successfully reproduces the average thermodynamic
stratification of smaller scale, more realistic numerical models where
the frequency-dependent radiative transfer equation in LTE is solved in
detail.  This technique retains the computational efficiency of earlier
parameterized methods, but does not require continual calibration against
more realistic simulations.  We find that with the new method we are 
able to initiate and sustain a stable convection pattern with a 
distribution of cell sizes and turnover times characteristic of solar 
granulation.  

The method presents a middle ground between realistic radiative MHD
models that solve the transfer equation in detail, and idealized 
models that simply impose a thermodynamic stratification, or ignore
the physics of radiative transport entirely.  The motivation for
developing this technique is to make feasible physics-based 
large-scale or global parameter space studies of the interaction of 
active region-scale magnetic fields with the small scale fields associated 
with granular convection in a domain that includes both a convection 
zone and corona.

Whether the approximate treatment captures enough of the essential
physics of the system still remains to be seen.  The technique is 
certainly limited by the fact that sideways transport is ignored,
and important physics of the chromosphere has yet to be included
in the current models.  Even so, we are able to generate solar-like 
convective turbulence in a physically self-consistent way, and follow 
the magnetic evolution of structures that thread the interface between 
the convective interior and corona.  

In particular, we presented two simulations that confirm the existence 
of a ``separatrix'' in the flow of magnetic energy from the interior 
to the atmosphere (see \opencite{Steiner2008}), and demonstrate that it 
is the mechanical energy of surface convection driven by strong radiative 
cooling that is the source of energy for this divergent flux of 
electromagnetic energy.  In a quiescent region, in the absence of 
strong concentrations of magnetic flux, our models suggest that it is 
the low photosphere that provides the source of electromagnetic energy to 
the chromosphere and corona, not the sub-surface layers.\\

\noindent {\bf Acknowledgments:} $\,$  This research was funded in part by 
the NASA Heliophysics Theory Program (grant NNX08AI56G), the NASA 
Living-With-a-Star TR\&T Program (grant NNX08AQ30G), and the NSF's AGS 
Program (grant ATM-0737836).  The authors wish to acknowledge Dick 
Canfield's pioneering efforts in the development of radiation hydrodynamics 
during the 1970s and 1980s. The particular description in this paper for 
the simplified radiative cooling treatment was inspired by the very first 
work (unpublished) that GHF did for Dick Canfield as 
a graduate student, namely an investigation of escape probability treatments 
for continuum radiation processes. Indeed, it is possible to derive the 
radiative cooling treatment described here using escape probability 
concepts. 

   
\bibliographystyle{spr-mp-sola}

\bibliography{papers}  

\end{article} 
\end{document}